# The Effect of Water on Quinone Redox Mediators in "Non"-aqueous Li-O₂ Batteries

Tao Liu,[a] James T. Frith,[b] Gunwoo Kim,[a,c] Rachel N. Kerber,[a] Nicolas Dubouis,[a]† Yuanlong Shao,[c] Zigeng Liu,[a]‡ Pieter M. Magusin,[a] Michael T.L. Casford,[a] Nuria Garcia-Araez,[b] Clare P. Grey[a]*

AUTHOR ADDRESS

a) Chemistry Department, University of Cambridge, Lensfield Road, Cambridge, UK CB2 1EW;

b) Chemistry Department, University of Southampton, Highfield Campus, Southampton, UK SO17 1BJ;

c) Cambridge Graphene Center, University of Cambridge, 9 JJ Thomson Avenue, Cambridge, UK CB3 0FA.

KEYWORDS ("lithium oxygen batteries, benzoquinone, hydrogen-bonding, redox mediators, $Li_2O_2$").

**ABSTRACT:** The parasitic reactions associated with reduced oxygen species and the difficulty in achieving the high theoretical capacity have been major issues plaguing development of practical non-aqueous Li-O₂ batteries. We hereby address the above issues by exploring the synergistic effect of 2,5-di-*tert*-butyl-1,4-benzoquinone and $H_2O$ on the oxygen chemistry in a non-aqueous Li-O₂ battery. Water stabilizes the quinone monoanion and dianion, shifting the reduction potentials of the quinone and monoanion to more positive values (vs. $Li^+$). When water and the quinone are used together in a (largely) non-aqueous Li-O₂ battery, the cell discharge operates via a two-electron oxygen reduction reaction to form $Li_2O_2$, the battery discharge voltage, rate, capacity all being considerably increased and fewer side reactions being detected; $Li_2O_2$ crystals can grow up to 30 μm, more than an order of magnitude larger than cases with the quinone alone or without any additives, suggesting that water is essential to promoting a solution dominated process with the quinone on discharging. The catalytic reduction of $O_2$ by the quinone monoanion is predominantly responsible for the attractive features mentioned above. Water stabilizes the quinone monoanion via hydrogen bond formation and by coordination of the $Li^+$ ions, and it also helps increase the solvation, concentration, life time and diffusion length of reduced oxygen species that dictate the discharge voltage, rate and capacity of the battery. When a redox mediator is also used to aid the charging process, a high-power, high energy-density, rechargeable Li-O₂ battery is obtained.

## 1. Introduction

Quinones represent an important class of organic redox molecules that are involved in energy transduction and storage in biological systems.[1-4] For example, they play a pivotal role in proton-coupled electron transfer for the natural respiratory and photosynthetic processes.[5] This unique charge transfer role of quinones inspired researchers to explore their applications in a range of artificial energy harvesting and storage devices, including dye-sensitized solar cells, artificial photosynthesis, pseudocapacitors, organic lithium ion batteries, redox flow batteries, and so on.[6-9] This versatility of quinones in part stems from the ease by which their physicochemical properties (redox, solubility, optical and electrical properties) can be tuned by engineering the molecular structures and through interactions with their chemical environment.[10-12] For example, the redox potentials of many quinone systems in nonaqueous media can be shifted to more positive values via the use of additives containing O-H and N-H bonds.[13-17] These observed shifts have been rationalized by hydrogen-bond formation of the negatively charged carbonyl oxygens on the reduced quinones with hydrogen atoms from water, alcohols or amines,[18-21] the nature of the hydrogen bond being characterized by electron spin resonance, ultraviolet-visible (UV-vis) spectroscopy and theoretical calculations.[22-24]

Recently, quinones have also been explored as redox mediators for the oxygen reduction reaction in Li-O₂ batteries. The non-aqueous Li-O₂ battery is considered as the ultimate battery as it possesses a theoretical energy density close to gasoline, 10 times higher than the state-of-art lithium ion battery.[25-28] Its operation typically involves $O_2$ reduction during discharge, the first step involving a one-electron electrochemical step to form $LiO_2$, which then chemically disproportionates to form $Li_2O_2$; a solid phase precipitates out of the liquid electrolyte and deposits on the porous electrode. On charging, the solid discharge product is decomposed releasing $O_2$. Realizing the theoretical capacity is, however, associated with significant challenges, in part because the electronically insulating discharge product tends to form as small particles or conformal films that quickly passivate electrode surfaces,[29-31] impeding further interfacial electron transfer and ion diffusion through the porous electrode. As a result, the cell discharge tends to finish early before the discharge product fully takes up the free volume available in the porous electrode. Furthermore, electrolyte decomposition promoted by reactions with reduced oxygen species,[32-42] particularly $LiO_2$, occurs during discharge. Therefore, enabling a mechanism that minimizes surface passivation, greatly promotes crystal growth of the discharge product and reduces the amount of side reactions, is key to realizing the full potential of a Li-O₂ battery.

Several redox couples,[43-48] including viologens,[43-44] phthalocranines[45] and quinones,[46-47] have been used to address the issues associated with the discharge of non-aqueous Li-O₂ batteries. A common feature is that these soluble molecules are able to chemically reduce $O_2$ in solution. The formation of surface passivation films is inhibited (to some extent) and the discharge capacity is increased. Another important aspect con-cerns their potential ability to reduce the life time[43-44] and decrease the free energies[47] of the chemically aggressive reaction intermediates, so that fewer side reactions occur.

In this work, we take advantage of the hydrogen-bonding properties of quinones[13-21] and evaluate the impact of the use of water with the quinone, 2,5-di-*tert*-butyl-1,4-benzoquinone (DBBQ), on the battery rate, capacity and side reactions. We show that the interactions with water stabilize the quinone monoanion, dianion and reduced oxygen intermediate species, the discharge becoming dominated by solution-phase processes. As a result, the rate and capacity are significantly improved. Importantly, the extent of side reactions was decreased by around 70%. We propose potential mechanisms with the aid of supporting experimental measurements and DFT calculations and discuss the significance of the work.

## 2. Results and Discussion

**Stabilization of DBBQ Monoanion and Dianion by Water.** The effect of added water on the redox chemistry of DBBQ is first evaluated. Figure 1 shows the linear voltage sweep experiments of DBBQ as a function of the water content in the LiTFSI/DME electrolyte. Under nominally dry conditions (<10 ppm water), two reduction peaks of DBBQ were observed, one at 2.5 V and the other at 2.15 V. These peaks are attributed to sequential one-electron reductions of the quinone (Q) to the quinone monoanion (Q$^{\bullet-}$) (I) and then to the quinone dianion (Q$^{2-}$) (II), each DBBQ molecule eventually taking up two electrons. This type of redox behavior is commonly observed for quinones in non-aqueous media.[15,18] As the water concentration was increased, the Q$^{2-}$ peak shifted positively by up to +0.4 V, whereas the Q$^{\bullet-}$ peak position shifted only slightly by +0.05 V. Above 20,000 ppm (1 M) $H_2O$ content, the quinone monoanion and dianion peaks merged together into a single peak, the peak area being the sum of those measured under anhydrous conditions (Figure S1 details the quantitative analysis.). A similar behavior was observed in an electrolyte with diglyme as the solvent and with another hydrogen-bonding donor molecule, methanol (Figure S1).

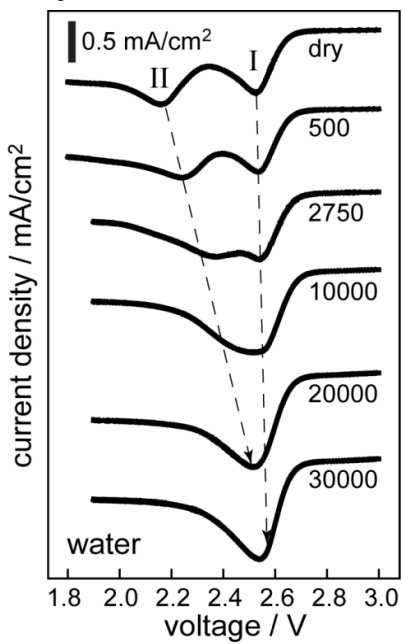

**Figure 1.** Linear voltage sweep measurements in a three-electrode setup of a cell as a function of the water content (in ppm, as labelled for each curve) in a 10 mM DBBQ, 0.25 M LiTFSI/DME electrolyte. A gas diffusion layer (GDL) electrode was used as the working electrode and the sweep rate is 10 mV/s for all experiments. Because of the high-water contents used, lithium iron phosphate was used as the reference and counter electrodes. All potentials are referenced against Li/Li$^+$ (-3.04 V versus standard hydrogen electrode).

To investigate the effect of water on the quinone formation further, we measured the UV-vis spectra of the quinones in anhydrous and wet conditions (Figure 2). In the anhydrous case, the cell was potentiostatically discharged at 2.5 V and 1.9 V until the current dropped to approximately zero; electrolyte samples at the corresponding voltages were extracted and subjected to UV-vis measurements to obtain reference UV-vis spectra of DBBQ, its anion and dianion (Figure 2A). Optically, the electrolyte underwent color changes from green (Q), to brown (Q$^{\bullet-}$) and light pink (Q$^{2-}$). The quinone UV-vis spectrum exhibits a major absorption at 255 nm and a weaker peak at 305 nm. For the quinone anion, in addition to the peaks at 255 and 305 nm, the spectrum also shows a broad absorption from 350 to 450 nm, and peaks at 235, 315, and 325 nm. The dianion spectrum has very distinct features: the absorption at 320 nm becomes the most intense peak, followed by the one at 240 nm and a broad absorption at around 400 nm.

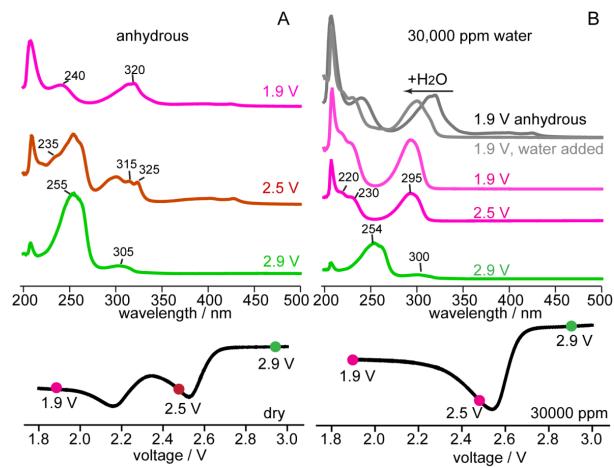

**Figure 2.** UV-vis spectra of DBBQ at different states of oxidation in an (A) anhydrous and (B) 3% $H_2O$ (30,000 ppm) added 0.25 M LiTFSI/DME electrolyte. The monoanion and dianions were obtained by potentiostatically reducing the quinone at 2.5 and 1.9 V, respectively; their corresponding spectra are color coded, that is, green for quinone, brown for semiquinone and pink for the dianion. For the case with added water, the quinone species generated potentiostatically at 2.5 and 1.9 V are presented with similar color codes. The effect of added water on the UV-vis spectrum of the dianion was investigated by adding an equivalent amount of $H_2O$ (3%) to the dianion sample obtained in the anhydrous electrolyte (grey). The sharp peak at 208 nm is very close to the spectrum cut-off and does not shift on addition of water. Its origin is unclear.

Moving to the case with 30,000 ppm water, the quinone spectrum shows absorption peaks at 254 and 300 nm, both slightly blue-shifted compared to that in the absence of water. On potentiostatically discharging the cell to only 2.5 V, the electrolyte became light pink in color indicating that the dianion is present. The capacity recorded was 90% of the theoretical value expected for a 2 electron per quinone molecule and further discharge at 1.9 V led to negligible capacity increase, confirming that almost all of the quinone had been reduced to the dianion at 2.5 V. The corresponding UV-vis spectra further support this view as the spectra (Figure 2B) at 2.5 and 1.9 V are nearly identical, showing a major peak at 295 nm and two weaker absorptions at 220 and 230 nm.

Compared to the anhydrous dianion spectrum (Figure 2B), the weak broad absorption at 400 nm was suppressed and the other peaks were greatly blue-shifted in the presence of water. The water-induced blue shifts were confirmed by adding water to the anhydrous dianion sample (Figure 2B, grey spectra); these shifts and the positive shifts of the half-wave potentials of reduced DBBQ due to added water are consistent with the hydrogen-bonding effects reported for many other quinones in nonaqueous media.[13,14,18,19,21,24]

Figure 3 shows the density functional theory (DFT) calculations for the Gibbs free energy of formation of the monoanion and the dianion in the presence of increasing numbers of coordinating water molecules. These calculations indicate that with increasing water concentrations, the energies of formation become more negative (more favorable), consistent with the positively shifted reduction potentials. Moreover, as illustrated in structures shown in Figure 3 (A and B), this increased thermodynamic stability of reduced quinine anions appears to result from hydrogen-bond formation between the negatively charged monoanion/dianion oxygens and the protons of water, hydrogen-bonding between water molecules (black dashed lines) and the coordination of the lithium ions by water. Because of the higher negative charge density in $Q^{2-}$ versus $Q^{\bullet-}$ and the presence of two $Li^+$ ions per anion, stronger interactions with water and thus larger positive potential shifts are usually observed for the dianion than for the monoanion[15,18].

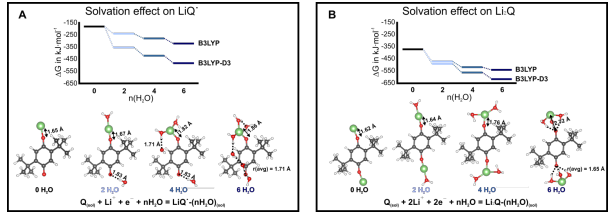

**Figure 3.** DFT calculations for the Gibbs free energy of formation for the monoanion $LiQ^{\bullet}$ (a) and the dianion $Li_2Q$ (b) in the presence of 0, 2, 4 and 6 coordinating water molecules. 20,000 ppm water in the electrolyte, as used experimentally in this work, corresponds to a water/$Li^+$ molar ratio of 4. The relevant energy minimized structures that result from these calculations are shown below, where the number of coordinating water used in the models is color coded from light to dark blue with the increasing $H_2O$ content. The green, dark grey, red and white balls correspond to Li, C, O, and H, respectively. The hydrogen bonds are illustrated with black dashed lines. In the absence of water, the $Li^+$ ions strongly coordinate to the oxygen atoms ($O_Q$) of the reduced quinone anions. The addition of water results in a gradual elongation of the r(Li-$O_Q$) bond, and ultimately, in the case of the dianion with six $H_2O$ molecules, complete shielding of the $Li^+$ from $Q^{2-}$ occurs, each $Li^+$ ion being solvated by three water molecules. The effect of the dispersion force correction (Grimme's dispersion with the original D3 damping function, see Experimental Details) is stronger for the case of the monoanion than for the dianion structural models, which is ascribed to the presence of more extensive hydrogen-bonding (H-O) interactions in the monoanion models.

**Improved Rate and Capacity**. Having established the effect of water on the redox chemistry of DBBQ, we next explore the role of water and DBBQ in a Li-O$_2$ battery. Figure 4A compares the linear voltage sweeps of cells with different electrolytes and atmospheres. To ensure a consistent electrode surface area for all tests, commercial gas diffusion layers (GDL) were used as the working electrode for all cells. Without DBBQ (blue curve), the oxygen reduction reaction (ORR) shows a peak current of -0.8 mA/cm$^2$, with an onset potential at 2.6 V. When only DBBQ was added to the electrolyte (red curve), both reductions peaks associated with the quinone monoanion ($Q^{\bullet-}$) and dianion ($Q^{2-}$) increased in the presence of O$_2$, the former (-2 mA/cm$^2$) being around 4 times higher in O$_2$ than in Ar; these observations suggest that the quinone monoanion catalyzes the oxygen reduction reaction, consistent with previous reports.[46-47] Adding a water content of 20,000 ppm (black curve), resulted in a greatly enhanced reduction current, the corresponding reduction peak (-5.2 mA/cm$^2$) being around an order of magnitude higher than that without DBBQ and water (blue curve), and increased by more than 2.5 times than that observed when using DBBQ alone (red curve). Furthermore, the onset voltage for reduction current is increased from 2.70 V to 2.85 V, consistent with the positively shifted reduction potential of DBBQ (Figure 1) due to water. Notably, the aforementioned effects cannot be solely due to added water, because the addition of water only (and with no DBBQ) only leads to a peak current of -2.2 mA/cm$^2$ and there is no shift in the onset reduction potential (green curve).

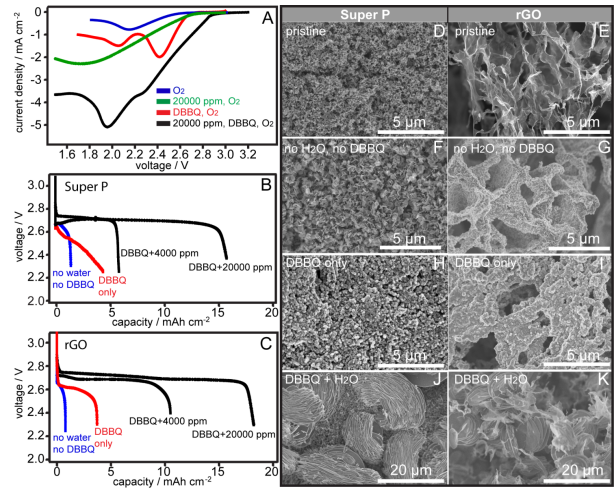

**Figure 4**. Electrochemistry of Li-O$_2$ batteries with different electrode structures and electrolytes (A-C) and SEM characterization of discharged electrodes (D-K). Linear voltage sweep measurements (A) of cells using GDL working electrodes with different electrolytes under O$_2$ (as labelled: water content in ppm and 10 mM for DBBQ concentration). Galvanostatic discharge curves of cells made of super P (B) and reduced graphene oxide (C) electrodes, either with neat 0.25 M LiTFSI/DME electrolyte, with only DBBQ added or with both DBBQ and H$_2$O added to the neat electrolyte (as labelled in the figures). D and E represent pristine super P and rGO electrodes. F (G), H (I), J (K) respectively represent super P (rGO) electrodes discharged in an anhydrous neat electrolyte, a neat electrolyte with 10 mM DBBQ, and an electrolyte with both 10 mM DBBQ and 20,000 ppm water added. All cells in (B and C) were discharged at 0.1 mA/cm$^2$; 15 mAh/cm$^2$ for SP and rGO electrodes is equivalent to 15,000 mAh/g$_c$ and 150,000 mAh/g$_c$, respectively.

To evaluate the effect of wet DBBQ on the discharge capacity, Li-O$_2$ batteries made of super P carbon and reduced graphene oxide electrodes were investigated. Figure 4(B-C) show the galvanostatic discharge curves of the Li-O$_2$ batteries with different electrolyte compositions. For both super P and graphene electrodes, the combined use of DBBQ with water leads to large capacity increases (up to 40 times with 20,000 ppm $H_2O$) compared with those of using neat electrolyte or only DBBQ; the respective discharge plateaus also shift to higher voltages, consistent with positively shifted reduction potentials of DBBQ in the presence of water (Figure 1 and 4A). The corresponding SEM images (Figure 4(D-K)) of the discharged electrodes reveal that in the neat electrolyte, the discharge product forms as small disc-like particles (F) or conformal films (G) covering the electrode surface. This is consistent with previous results obtained in weak solvating electrolytes, where $Li_2O_2$ formation is dominated by a surface mechanism.[29-30] With added DBBQ, thicker layers of agglomerates comprising small particles form on the discharged super P (H) and rGO (I) surfaces. In the case of DBBQ and water, large particles of 10-20 μm were observed for both discharged super P (J) and rGO electrodes (K) and bare super P and rGO electrode surfaces can be still seen (more SEM images in Figure S2); this result supports an enhanced solution mediated mechanism during discharge, which accounts for the large capacities observed. Similar phenomena were observed with GDL electrodes, as used in a previous study.[47] The discharge capacity was considerably increased when DBBQ was used together with water, compared to DBBQ alone; larger discharge particles and agglomerates of particles were formed in the discharged GDL electrodes (Figure S3). Of note, in the macroporous graphene electrode (H) there was still plenty of open space for the particles to grow in contrast to the super P electrode (G); the termination of discharge for the rGO battery was instead limited by the loading of the $LiFePO_4$ (LFP) counter electrode (around 80 mg), used so as to avoid problems with lithium metal in the presence of excess water. Compared with a mesoporous structure, a macroporous electrode structure enables the growth of larger crystals and appears to allow unhindered diffusion of redox species even at deep discharges.

**A $Li_2O_2$ Chemistry, But with Fewer Side Reactions.** Because of the large amount of water used, it is important to verify the chemical nature of the discharge product. XRD measurements for discharged super P electrodes (Figure 5A) confirm that $Li_2O_2$ is the only crystalline discharge product in DBBQ electrolytes with and without added water – essentially no crystalline LiOH is formed;[31,49] the former case leads to much higher $Li_2O_2$ crystallinity, as also evidenced by SEM (Figure 4J). $^7Li$ solid state NMR measurements (Figure 5B) show a single resonance at around 0.35 ppm, suggesting that $Li_2O_2$ is the dominant discharge product.[31,50] Differential electrochemical mass spectrometry (DEMS) experiments were also conducted to confirm the $e^-/O_2$ molar ratio during the course of cycling. It can be seen for both cases that during discharge, the $e^-/O_2$ molar ratio stayed close to 2 and no obvious $CO_2$ or $H_2$ evolution was observed; this suggests a process that consumes 2 electrons per $O_2$ reduction and is consistent with $Li_2O_2$ formation. *Operando* pressure measurements performed on discharging also strongly support that the quinone mediated oxygen reduction reaction closely follows $2e^-$ per reacting $O_2$, at rates up to 1 mA/cm² (Figure S4). To determine the extent of any side reactions of the quinone-involved chemistry, we performed quantitative $^1H$ magic angle spinning NMR measurements by using a spin flip angle of 30° and a sufficiently long recycle delay of 200 s. The $^1H$ spectrum (Figure 5C) for the neat electrolyte exhibits resonances at 1.9, 3.5, 8 and -1.5 ppm, which represent the typical side reaction products of lithium acetate, methoxide, formate and hydroxide, respectively, in an ether based electrolyte. These side reactions are mainly caused by nucleophilic attack of ether by the superoxide and peroxide.[31-38] The use of DBBQ led to little change in the nature or the quantity of the side reaction products. In the presence of both DBBQ and 20,000 ppm water, however, around 70% reduction in the $^1H$ signal was observed, demonstrating that all the aforementioned side reaction products were reduced; this conclusion is consistent with a corresponding narrower $^7Li$ NMR resonance centered at 0.35 ppm, i.e., fewer types of lithium-containing chemical environments are present (Figure 5B).

The rechargeability of the battery was also assessed (Figure S5), charging being associated with a plateau at 3.6 V. In the subsequent cycling, however, the 3.6 V plateau shortens and an additional charging plateau at 4.2 V appeared that dominated the charging process, the Coulombic efficiency decreases steadily on cycling. A high voltage plateau (above 4.0 V) is typically observed in anhydrous $Li_2O_2$ based $Li-O_2$ batteries. To explore this further, DEMS measurements were performed on charging: Figure 5(F, G) compares gas evolutions following a discharge (as presented in Figure 5(D, E)), respectively. For the case with DBBQ alone, the battery can only recharge to around 1/3 of the prior discharge capacity (i.e., 33% coulombic efficiency) under the conditions used in the DEMS set-up, with the charging voltage rapidly rising from 3.4 to 4.5 V. The corresponding $O_2$ signal (F) is considerably less than that expected for a process involving 2 electrons per $O_2$ evolved, and there is an accompanying $CO_2$ evolution, suggesting that side reactions occur on charging. The negative $H_2$ intensity is a result of a decrease in the $H_2$ background signal and then a baseline correction of the data; this $H_2$ intensity decrease could also be due to side reactions that consume electrolyte components contributing to the $H_2$ background signal (e.g., from diglyme and $H_2O$). On the other hand, the case with a combined use of DBBQ and water is associated with a rapidly increasing voltage until a charging plateau at 4.2 V (G) is observed, the recharging finishing with 69% coulombic efficiency at 4.5 V. The corresponding $O_2$ signal on charging rose to the expected level of 2 electrons per $O_2$ evolved. It then dropped down by half and afterwards gradually climbed up again to the expected level. It is clear in this case that the $e^-/O_2$ molar ratio stays closer to 2, compared to the case without water, i.e., the rechargeability is improved with water. The dip in the $O_2$ DEMS signal at the beginning of charge is tentatively ascribed to a competing water oxidation reaction, which involves 4 electrons per $O_2$ evolved, and thus a decrease in the $O_2$ evolution signal. This $O_2$ signal decrease coincides with degradation reactions evolving $H_2$. As the water oxidation reactions and other side reactions involving $CO_2$ evolution (G) continued, the water content was reduced and/or the carbon surface became passivated, which slowly shifts the charging reaction closer to 2 electrons per $O_2$ evolution, that ratio expected for $Li_2O_2$ decomposition. These observations are consistent with our proposal that the gradual increase in the cell charging voltages (Figure S5) to potentials similar to those seen in the absence of added water are due to the electrochemical loss of water in the electrolyte in the DEMS cell.

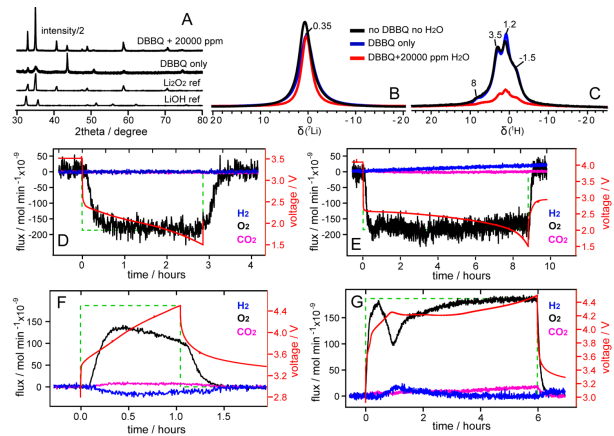

**Figure 5**. XRD (A), solid state $^7Li$ and $^1H$ MAS NMR spectra of discharged electrodes (B, C) and *operando* DEMS measurements (D-E) of the $Li-O_2$ battery system. (A) The XRD patterns of electrodes fully

discharged in anhydrous and wet (20,000 ppm H$_2$O) 0.01 M DBBQ 0.25 M LiTFSI/DME electrolyte (as labelled); the XRD patterns of Li$_2$O$_2$ and LiOH reference powders are plotted for comparison. $^7$Li NMR (B) and quantitative $^1$H NMR spectra (C) spectra comparing the amount of side reactions involved in cells made of super P electrodes and different electrolytes (0.25 M LiTFSI/DME without any additive, with only DBBQ and with DBBQ and 20,000 ppm water, as labeled). DEMS signals measured at a continuous flow mode for cells with a nominally anhydrous (D-discharging, F-charging) and 2% H$_2$O added (E-discharging, G-charging) electrolytes (0.01 M DBBQ 0.25 M LiTFSI/diglyme). Diglyme was used because DME evaporates too fast in a continuous flow mode, even at a slow flow rate of 80 μl/min. Red, black, blue and pink colors represent the cell electrochemistry, O$_2$, H$_2$ and CO$_2$ mass spectrometry signals, respectively. The green broken lines show the oxygen signals expected for an ideal 2 electrons per O$_2$ consumption/evolution.

The water oxidation issue can be resolved by introducing the soluble charging mediator 2,2,6,6-tetramethyl-1-piperidinyloxy (TEMPO) together with DBBQ (Figure 6). TEMPO was chosen since it has previously been demonstrated as an effective charging mediator in an ether electrolyte.[51] By using just 20 mM TEMPO, the charging voltage plateau (A) decreases to around 3.6 V; *operando* pressure measurements confirm oxygen evolution on charging the battery (Figure S6). At 1 mA/cm$^2$, the cell still has a capacity of more than 4 mAh/cm$^2$ (or 4000 mAh/g$_c$), whereas without DBBQ and TEMPO, the cell has very little capacity on charging even at a lower rate of 0.5 mA/cm$^2$, the voltage rapidly rising to the cut-off value of 4.5 V. When the discharge capacity is curtailed at 1 mAh/cm$^2$ (or 1000 mAh/g$_c$) (Figure 6B), the cell cycles much more stably (as compared to the case without TEMPO). It is clear that by using DBBQ, TEMPO and water, the energy efficiency and rechargeability of the battery have been improved.

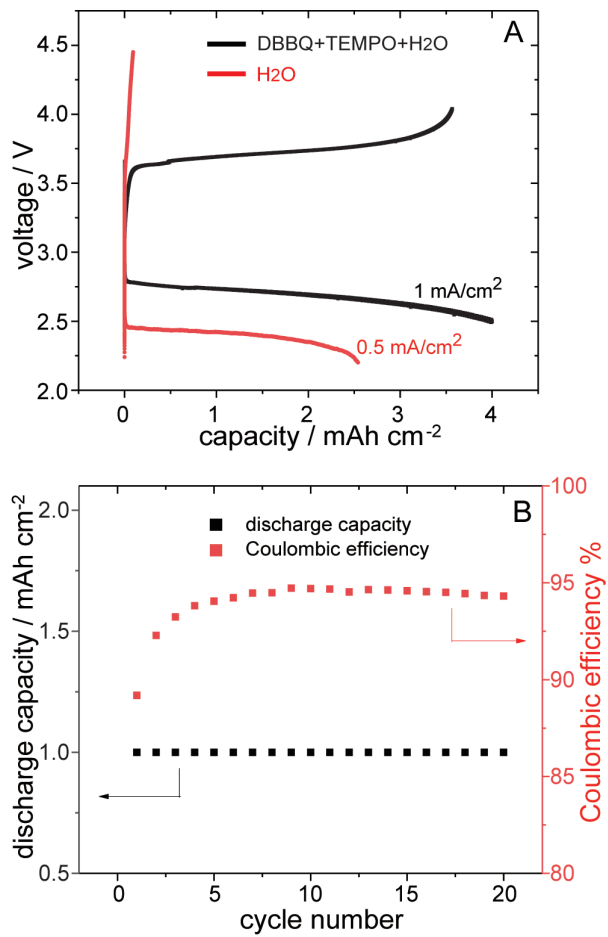

**Figure 6.** Electrochemistry of Li-O$_2$ batteries made using a 0.25 M LiTFSI/diglyme electrolyte with DBBQ (10 mM), TEMPO (20 mM) mediators and 20,000 ppm added water. Cells cycled at different rates (up to 1 mA/cm$^2$) and at different depth of discharge are shown in (A); (B) shows the cycling stability of a cell in (A) at 0.5 mA/cm$^2$, capacity limited at 1 mAh/cm$^2$. Carbonized polyacrylonitrile electrodes were used here, where the macroporous electrode structure facilitates diffusion of redox species at higher rates.

In summary, we have shown that with a combined use of DBBQ and water, the discharge rate, capacity and overpotentials of the Li-O$_2$ battery are all improved. The oxygen reduction becomes dominated by a solution phase process. When a charging mediator is also added, the rechargeability of the battery is further improved. This opens up the possibility of developing a more efficient, high power Li-O$_2$ battery by choosing alternative charging mediators with lower redox potentials than TEMPO and by further optimizing the mediator concentrations. Importantly, fewer side reactions are involved during both cell discharge and charge in the battery. In the next section, we discuss the potential mechanisms responsible for the attractive features of the current oxygen chemistry.

**Mechanism Interpretation.** To enable a fast-rate, solution dominated discharge process with fewer side reactions in lithium oxygen batteries, the following factors are important: (1) the concentration of dissolved O$_2$ in the electrolyte, (2) the concentration and the thermodynamic stability of the reaction intermediates (which control their life time, diffusion length in the electrolyte and chemical reactivity), and (3) the solubility and thermodynamic stability of the final discharge product(s). The ensuing discussion focuses on the impacts of water on the above aspects in the current DBBQ-mediated Li-O$_2$ battery system.

In the *anhydrous* case, the galvanostatic discharge (Figure 4 B, C) occurs between 2.3 and 2.6 V, where only the quinone monoanion can be formed according to the CV obtained under anhydrous conditions (Figure 1). Since the reduction potential of Q/Q$^{\bullet-}$ is higher than that of O$_2$/O$_2^-$ (Figures 1 and 2), the monoanion will necessarily be generated prior to the formation of superoxide anions during discharging. Previous studies[36-37] suggested that the quinone monoanion can chemically reduce O$_2$, where the formation of Li$_2$O$_2$ was confirmed in the solid precipitate after mixing the monoanion with O$_2$. To verify the proposed reactions of the monoanion Q$^{\bullet-}$ with O$_2$, the reduced quinone species generated in anhydrous conditions was exposed to pure O$_2$; the corresponding UV-vis spectrum (Figure 7) shows that the quinone monoanion can indeed chemically reduce O$_2$, itself being re-oxidized to quinone again. The increased oxygen reduction current (Figure 4A, at above 2.3 V) is hence attributed to the catalysis of O$_2$ by the electrochemically generated monoanion.

The possible elementary steps are summarized in Reactions 1–6.

(1) Q$_{(sol)}$ + Li$^+$ + e$^-$ → LiQ$^{\bullet}_{(sol)}$;

(2) LiQ$^{\bullet}_{(sol)}$ + O$_{2(sol)}$ → LiQO$_{2(sol)}$;

(3) 2LiQO$_{2(sol)}$ → Li$_2$O$_{2(s)}$ + O$_{2(sol)}$ + 2Q$_{(sol)}$;

(4) LiQO$_{2(sol)}$ + LiQ$^{\bullet}_{(sol)}$ → Li$_2$O$_{2(s)}$ + 2Q$_{(sol)}$;

(5) LiQO$_{2(sol)}$ → LiO$_{2(sol)}$ + Q$_{(sol)}$;

(6) 2LiO$_{2(sol)}$ → Li$_2$O$_{2(s)}$ + O$_{2(sol)}$.

In the absence of water, a quinone molecule is first electrochemically reduced to form a lithium quinone monoanion, LiQ$^{\bullet}$, as a soluble solvated species (sol) in the electrolyte (Reaction 1). The diffusion length of the soluble monoanion away from the surface can potentially be very long, until it meets an O$_2$ molecule and chemically reduces it. This chemical reduction of O$_2$ to Li$_2$O$_2$ could proceed via two pathways. One involves LiQ$^{\bullet}_{(sol)}$ reacting with O$_2$ to form LiQO$_{2(sol)}$; two LiQO$_{2(sol)}$ then disproportionate to form Li$_2$O$_2$ (Reactions 2-3), or LiQO$_{2(sol)}$ reacts with LiQ$^{\bullet}_{(sol)}$ to form Li$_2$O$_2$ (Reaction 4). The other path involves the dissociation of LiQO$_{2(sol)}$ to form LiO$_{2(sol)}$ and Q$_{(sol)}$, the former further decomposing to form Li$_2$O$_2$ (Reactions 5-6). In a previous study, Gao et al.[47] proposed a very similar mechanism as described in Reactions (1-4). It was further suggested that the reactive LiO$_2$ species was circumvented, i.e., LiQO$_{2(sol)}$ does not dissociate into LiO$_2$ and Q, which was in part supported by the observation of a small reduction in the concentration of parasitic reaction products in their batteries: 4% and 5% fewer side reaction products were seen in DME and TEGDME electrolytes, respectively with DBBQ than without it. In our experiments (Figure 5C), the quantities of degradation products formed in the DME electrolytes with or without DBBQ were extremely similar so that the little difference is within the error limit of our quantification experiments. Given that no substantial reduction of the side reactions was seen with the use of DBBQ, we suggest that Reactions 4 and 5 *are* likely to occur in addition to the disproportionation (Reaction 3). In other words, LiO$_2$ still exists as an intermediate species, being responsible for a significant amount of side reactions observed during discharge. In the anhydrous electrolyte with DBBQ, LiQO$_{2(sol)}$ complex formation and the high solubility of LiQ$^{\bullet}_{(sol)}$, to some extent, help to generate/transport reduced oxygen species (LiO$_2$ and Li$_2$O$_2$) further away from the surface, promoting the formation of thicker layers of Li$_2$O$_2$ as compared to the case without DBBQ, as illustrated in Figure 4 (F-I). From the SEM images (Figure 4 H, I), the discharge process is still largely confined to regions near electrode surfaces (within 200 nm); this observation supports the view that LiQ$_{(sol)}$ is chemically oxidized by nearby O$_{2(sol)}$ before it diffuses further into the bulk electrolyte and that the complex LiQO$_{2(sol)}$ and Li$_2$O$_{2(sol)}$ do not diffuse very far into the electrolyte either.

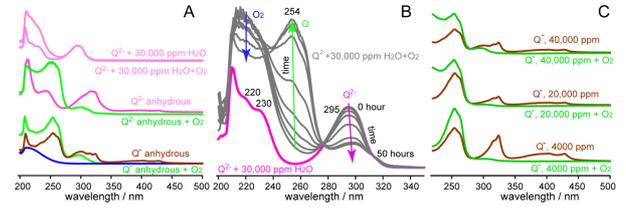

**Figure 7**. UV-vis spectra performed to determine the abilities of the quinone monoanion and dianion to chemically reduce O$_2$ at dry and wet conditions. UV-vis spectra of the monoanion (Q$^{\bullet-}$), dianion (Q$^{2-}$) generated under anhydrous conditions (0.25 M LiTFSI/DME, 0.01 M DBBQ) and the dianion generated under wet conditions (30,000 ppm H$_2$O) (A), after exposing them to an excess of pure O$_2$; the spectra prior to O$_2$ exposure are plotted for comparison. The time-dependent reaction of the quinone dianion with O$_2$ under wet conditions (30,000 ppm H$_2$O) is shown in B. UV-vis spectra of quinone monoanions mixed with wet electrolytes (4000, 20,000 and 40,000 ppm water) and then exposed to O$_2$ (C). The anhydrous monoanion and dianion react immediately to reform the quinone (A). In the wet case, the dianion spectrum remained unchanged after O$_2$ exposure (A and B), the extent of reaction only becoming appreciable after many hours, the absorption (254 nm) due to Q gradually increasing whilst those associated with O$_2$ and Q$^{2-}$ (295 nm) decrease (arrows in B). The general increase in the spectral background in the UV region is due to absorption of excess O$_2$ (blue curve in A, obtained by exposing an anhydrous DME solvent to O$_2$). Quinone monoanion (Q$^{\bullet-}$) in the presence of water (4000 to 40,000 ppm) can still readily react with O$_2$, itself being oxidized back to the quinone state (Q).

Turning to the *wet* case with 20,000 ppm water, the redox chemistry of hydroquinones needs to be taken into consideration, because it is well known that many hydroquinone monoanions are able to reduce oxygen.[16-17] Indeed, millions of tons of H$_2$O$_2$ are produced annually via the oxidation of hydroantraquinone by O$_2$.[52] One thus needs to consider whether in the presence of such a large amount of H$_2$O (H$_2$O to DBBQ molar ratio being equal to 100:1 at 20,000 ppm), the oxygen reduction reaction in the current system is actually mediated by the equivalent hydroquinone instead of DBBQ. This possibility was ruled out by performing CV measurements of the 2,5-di-*tert*-butyl-hydroquinone in the same water added DME electrolyte (Figure S7), which showed a redox potential at around 3.5 V, considerably higher than the redox process seen in Figure 4. Clearly the hydroquinone is not formed under the present conditions and it is not responsible for the oxygen chemistry observed in this work.

Indeed, Gupta and Linschitz demonstrated that the mechanism of reduction of quinones in wet organic solvents does not involve protonation of the quinone monoanion or dianion.[19] Protonation of the reduced quinones could be ruled out in their work, because it was found that their pKa's were much lower than those of water or alcohols, meaning that it was much easier to deprotonate reduced quinones than it is to extract a proton from water or alcohols. Therefore, the extraction of a proton from water by the quinone monoanion or dianion can be ruled out on thermodynamic grounds: i.e., Q$^{\bullet-}$ + H$_2$O = HQ$^{\bullet}$ + OH$^-$ is thermodynamically unfavourable, whereas stabilization of the quinone monoanion by hydrogen bonding with water is thermodynamically favourable, again as confirmed in DFT studies of small clusters (Figure 3).

In the presence of more than 20,000 ppm water (Figures 1 and 4), both the benzquinone monoanion and the dianion can be formed during the galvanostatic discharge at 2.3-2.8 V, and CV/DEMS/pressure measurements (Figures 4, 5 and S4) show that the reduced quinone definitely mediated the two-electron reduction reaction. To verify its reactivity with $O_2$, the dianion species generated in either an anhydrous or wet electrolyte were investigated with UV-vis spectroscopy after being exposed to an excess of $O_2$. As shown in Figure 7, although the dianion under anhydrous conditions can readily reduce $O_2$ (A), the reaction of quinone dianion with $O_2$ in the presence of 20,000 ppm water is sluggish. There was no appreciable change in the spectrum within an hour since the $O_2$ exposure, and there was still around 1/3 of the quinone dianion unreacted (Figure 7B) even after 50 hours; this observation is inconsistent with the fast DEMS response that supports two-electron ORR activity in Figures 5E and S4, suggesting that oxygen reduction mediated by the quinone dianion is unlikely to be the dominant process on discharging. On the other hand, the quinone monoanion in the presence water was found to react readily with oxygen in the presence of 4000 to 40,000 ppm water. Figure 7C shows that the characteristic absorption bands between 300 and 450 nm associated with the quinone monoanion disappeared after exposing the sample to $O_2$, and the resulting spectra support that the oxidized quinone state, Q, is formed. *In situ* Raman measurements further suggest that in the presence of 20,000 ppm water, the quinone monoanion is indeed an intermediate species during the reduction to form the dianion (Figure S8). We therefore ascribe the enhanced $O_2$ reduction at 20,000 ppm water to the catalysis of quinone monoanion.

The potential reaction pathways are summarized in Reactions (7-11). The quinone molecule is first electrochemically reduced to form the lithium quinone monoanion (Reaction 7); the monoanion then chemically reacts with dissolved $O_2$ to form a water-coordinated monoanion-oxygen complex, $LiQO_2\cdot(nH_2O)_{(sol)}$ (Reaction 8), where n (=1 to 4) is the number of the coordinating water with the complex. Subsequently, the monoanion-oxygen complex disproportionates or reacts with another quinone monoanion to form $Li_2O_2$, Reactions 9-11. Given the fewer side reactions observed in the presence of water (Figure 5B and C), we tentatively propose that the $LiO_2$ intermediate has been circumvented in the wet case.

(7) $Q_{(sol)} + Li^+_{(sol)} + e^- + nH_2O \rightarrow LiQ^\bullet\cdot(nH_2O)_{(sol)}$ 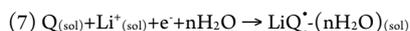

(8) $LiQ^\bullet\cdot(nH_2O)_{(sol)} + O_{2(sol)} \rightarrow LiQO_2\cdot(nH_2O)_{(sol)}$ 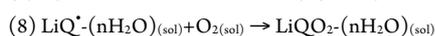

(9) $2LiQO_2\cdot(nH_2O)_{(sol)} \rightarrow Li_2O_{2(sol)} + 2Q + 2nH_2O + O_2$ 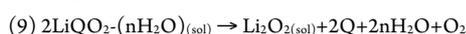

(10) $LiQO_2\cdot(nH_2O)_{(sol)} + LiQ^\bullet\cdot(nH_2O)_{(sol)} \rightarrow Li_2O_{2(sol)} + 2Q + 2nH_2O + O_2$ 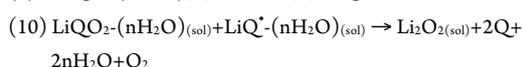

(11) $Li_2O_{2(sol)} \rightarrow Li_2O_{2(s)}$ 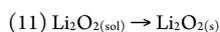

Despite the fact that the quinone dianion reacts slowly towards $O_2$ reduction, its formation is not detrimental for the battery operation, because it can react with the neutral quinone via the following comproportionation reaction (12)[53-54] to form the monoanion that readily reduces $O_2$. This hypothesis is consistent with the observation (Figure S4) that the battery can operate closely via two electrons per reduced $O_2$ at high rates.

(12) $Q_{(sol)} + Li_2Q\cdot(nH_2O)_{(sol)} \rightarrow LiQ^\bullet\cdot(nH_2O)_{(sol)}$ 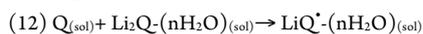

From the SEM images (Figure 4 J, K and Figure S2), the $Li_2O_2$ crystals grow as large as 30 μm, suggesting that reduced oxygen species, $Li_2O_{2(sol)}$ and/or $LiQO_2\cdot(nH_2O)_{(sol)}$, were generated at a location or had diffused by a distance up to 30 μm away from the electrode surface, much longer than that in the absence of water. The final product $Li_2O_2$ is much more soluble in water than in ether, the added water thus helping $Li_2O_2$ to diffuse further away from the electrode surfaces. However, cells discharged with 20,000 ppm added water but no DBBQ show that $Li_2O_2$ particles can only grow up to 3 μm at the end of the discharge (Figure S9), which implies that the water solvation alone cannot explain the phenomena seen with both DBBQ and water added. Given that added water tends to slow down the reaction of reduced quinone with $O_2$ (e.g. for the dianion), it is possible that in a $O_2$ saturated electrolyte, the water-coordinated monoanion can diffuse for a longer distance away from the electrode surfaces before it reduces $O_2$ (compared to the anhydrous case), effectively enlarging the reaction zone in the electrolyte and allowing reactions to occur deeper into the electrolyte. In addition, the water induced hydrogen-bond formation and $Li^+$ solvation via the quinone monoanion can help increase the concentration and diffusion length of $LiQO_2\cdot(nH_2O)_{(sol)}$ in the electrolyte. All these factors together increase the concentrations of the reduced oxygen reaction intermediates and helped promote a faster-rate, high-capacity discharge dominated by a solution phase process.

## 3. Conclusions

In conclusion, we have shown that the use of solvating additives with a H-bond formation ability is a powerful method to tune the thermodynamic stability of the reduced quinone species and its interactions with $O_2$, which in turn dictates the redox potential, the chemical reactivity, the solvation, the life time and diffusion length of these species in electrolytes. As a result, the discharge voltage, capacity and rate ability are all improved. Moreover, it helps enables a mechanism that is likely to help circumvent the formation of $LiO_2$, and thus a discharge process with fewer parasitic reactions. Together with a charging mediator, a more energy efficient, high power, rechargeable Li-$O_2$ battery is obtained. These effects of water are applicable to DBBQ in different aprotic solvents and hydrogen-bonding donors, and likely applicable to other soluble redox mediators that reduce $O_2$, increasing the options and likelihood of finding an optimal system compatible with a Li metal anode.

## AUTHOR INFORMATION

### Corresponding Author

* cpg27@cam.ac.uk.

### Present Addresses

† Chimie du Solide et de I'Energie, College de France, UMR 8260, 75231, Paris Cedex 05, France

‡ Forschungszentrum Juelich GmbH, Institute of Energy and Climate Research Fundamental Electrochemistry (IEK-9) 52425, Juelich and Max-Planck-Institute for Chemical Energy Conversion, Stiftstrasse 34-36, 45470, Muelheim an der Ruhr.

## ACKNOWLEDGMENT

The authors thank EPSRC-EP/M009521/1 (TL, GK, CPG), Innovate UK (TL), Darwin Schlumberger Fellowship (TL), EU Horizon 2020 GrapheneCore1-No.696656 (GK, CPG), EPSRC - EP/N024303/1, EP/L019469/1 (NGA, JTF), Royal Society - RG130523 (NGA) and the European Commission FP7-MC-CIG Funlab, 630162 (NGA) for research funding.

Table of Content

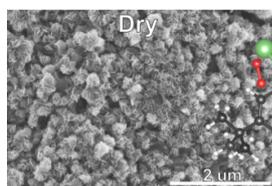 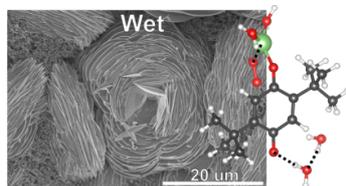

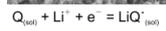 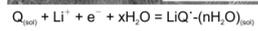

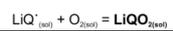 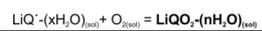